\documentstyle[11pt,fleqn]{article}
\newcommand{\Section}[1]%
{\section{#1}\setcounter{equation}{0}%
\setcounter{theorem}{0}}

{\par\noindent{\em #1:\ }}%
{~\rule{2mm}{2mm}\par\bigskip}
\newcommand{\ret}{\nonumber \\}
\newcommand{\beq}{\begin{equation}}
\newcommand{\eeq}{\end{equation}}
\newcommand{\beqar}{\begin{eqnarray}}
\newcommand{\eeqar}{\end{eqnarray}}

\newcommand{\abs}[1]{\left| #1 \right|}

\newcommand{\dagg}{\scriptscriptstyle\dagger}
\font\titlefnt=cmbx12 scaled \magstep2

\begin{document}
\mathindent 0mm
\setcounter{page}{0}
\newpage\thispagestyle{empty}
\topskip 3cm
\begin{center}
{\titlefnt Flow equations for the\\ \vspace*{0.4cm}
Anderson Hamiltonian}
\vskip3cm
Stefan~K.~Kehrein\footnote[1]{E--mail: ej6@vm.urz.uni-heidelberg.de}~~and
Andreas Mielke\footnote[2]{E--mail: mielke@hybrid.tphys.uni-heidelberg.de}
\\
\vspace*{0.2cm}
Institut f\"ur Theoretische Physik,\\
Ruprecht--Karls--Universit\"at,\\
D-69120~Heidelberg, F.R.~Germany
\\
\vspace*{1.5cm}
February 9, 1994
\\
\vspace*{1cm}
to appear in Journal of Physics A: Mathematical and General
\\
\vspace*{2cm}
{\bf Abstract}
\end{center}
Using a continuous unitary transformation recently proposed by Wegner
\cite{Wegner} together with an approximation that neglects irrelevant
contributions, we obtain flow equations for Hamiltonians. These flow
equations yield a diagonal or almost diagonal Hamiltonian.
As an example we investigate the Anderson Hamiltonian for dilute
magnetic alloys. We study the different fixed points
of the flow equations and the corresponding
relevant, marginal or irrelevant contributions. Our results are
consistent with results obtained by a numerical renormalization group
method, but our approach is considerably simpler.
\vspace*{3cm}
\pagebreak
\topskip 0cm
\Section{Introduction}
Recently Wegner \cite{Wegner} developed a metequations for Hamiltonians. These
equations result from a continuous
unitary transformation that brings the Hamiltonian closer to diagonalization.
The continuous unitary transformation generates a Hamiltonian $H(\ell)$ from
an initial Hamiltonian $H(0)$. It may be written in the form
\beq \label{flq1}
\frac{dH(\ell)}{d\ell}=[\eta(\ell),H(\ell)]
\eeq
where $\eta(\ell)$ is an antihermitian operator depending on $\ell$ as well.
Assume that $H$ is written in the form $H=H^d+H^r$. $H^d$ is a Hamiltonian
that can be diagonalized whereas $H^r$ contains further terms which are not
simple. We now try to choose $\eta$ such that $H^r(\ell)$ tends to zero
as $\ell$ goes to~$\infty$. Wegner proposed
\beq \label{eta}
\eta=[H,H^r].
\eeq
With this choice of $\eta$, $H^r$ does not necessarily vanish for
$\ell\rightarrow\infty$, but $\eta$ vanishes in this limit so that
$H$ can be diagonalized up to degeneracies of eigenvalues of $H^d$.
Notice that of course the method of unitary transformations of a
Hamiltonian is perfectly well known in solid state theory, in particular
to study non-perturbative effets. However, it is usually nontrivial to
find these unitary transformation. With Wegner's choice of $\eta$ in
(\ref{eta}) one has a general framework to construct such unitary
transformations that make Hamiltonians "simpler". One seems to recover
a lot of standard unitary transformations in this way. A well known
example is the Schrieffer-Wolff transformation \cite{SW}.
The connection between the unitary transformation induced by the
flow equations and the Schrieffer-Wolff transformation will be discussed
in section \ref{SRio}.

It is clear that it is in general impossible to solve
the full flow equations for a given initial Hamiltonian.
The situation is even worse. If we consider for example an initial
Hamilttwo-particle interaction, the flow equations generate additional
interactions among three and more particles. Wegner showed that in the
case of a $n$-orbital model in the limit $n\rightarrow\infty$, the
equations for the two-particle interaction are closed.
This fact allowed him to solve the flow equations for a one-dimensional
model in the limit $n\rightarrow\infty$. Then it is possible to perform
a $1/n$-expansion.

In the present paper we introduce an approximation to the flow
equations which allows to treat a large class of models. The approximation
consists of neglecting those contributions on the right hand side of
(\ref{flq1}) which are of a form different from the terms in the
initial Hamiltonian $H(0)$. E.g. the terms generating three-particle
interactions mentioned above are neglected. Then one obtains a set of
differential equations which can be analyzed.
In a second step one can now add different contributions to the Hamiltonian
in order to study the effect of the terms neglected before. If such a
contribution changes the Hamiltonian for large $\ell$ significantly, it
is relevant and must be included. If it does not change the Hamiltonian
for large $\ell$ it is irrelevant. As we will see, it is also possible that
such an additional contribution does not change the original terms in the
Hamiltonian but that it does not vanish either for large $\ell$. In this
case the contribution is called marginal. After having solved the
flow equations, the goal is to find all the relevant and marginal
contributions to the Hamiltonian.

In order to show how this method works, we investigate the Anderson model
for dilute magnetic alloys \cite{Anderson}.
It describes electrons on a lattice with
a single defect state. The Hamiltonian contains the kinetic energy of
thelectrons moving on the lattice, the energy of the defect state, a
hybridization of the states in the band with the defect state, and
a (usually repulsive) interaction of electrons in the defect state.
For an excellent overview on this model and the
results obtainable by different methods see \cite{Hewson}.
In a certain limit this model
is solvable using a Bethe ansatz. Furthermore, it has been investigated
by Krishnamurty, Wilkins and Wilson \cite{KWW} using a numerical
renormalization technique. We compare our results with some of the
results in \cite{KWW}. Especially we will see that the flow equations
yield "`fixed points"' similar to the fixed points found in \cite{KWW} and
that the operators which are relevant (marginal, irrelevant) in our
approach are relevant (marginal, irrelevant) in the renormalization group
approach as well.
We will come back to the connection to renormalization techniques
implied by our suggestive use of language later.

Our paper is organized as follows. In the next section we give a more
detailed description of the method. After that we derive the flow equations
for the Anderson model and we discuss the explicit structure
of these equations in detail.
The fixed points of the equations are described.
We also present numerical solutions of the equations.
In section \ref{SRio} we discuss possible relevant and irrelevant
contributions to the original Hamiltonian,
and in section \ref{SD} we compare our results for the
Anderson model with the results of \cite{KWW}.
Finally we summarize our results and briefly discuss possible further
investigations.

\Section{The method} \label{STm}
The main part of the method is already described by (\ref{flq1}) and
(\ref{eta}). It has been explained in detail by WBut since it is not yet well
known and since we introduce a new
approximation not used in \cite{Wegner}, we want to point out  some
useful properties of the flow equations. To do this, let $H=(h_{k,q})$
be a real, symmetric $N \times N$-matrix and $H^d$ its diagonal part.
$H^r$ contains the off-diagonal matrix elements.
Then the flow equations for $H(\ell)$ can be written in the form
\beq \label{flqm1}
\frac{dh_{k,q}(\ell)}{d\ell}=\sum_p(\eta_{k,p}(\ell)h_{p,q}(\ell)
-h_{k,p}(\ell)\eta_{p,q}(\ell))
\eeq
and
\beq \label{etam1}
\eta_{k,q}(\ell)=(h_{k,k}(\ell)-h_{q,q}(\ell))h_{k,q}(\ell)
\eeq
so that we obtain
\beq \label{flqm2}
\frac{dh_{k,q}(\ell)}{d\ell}=\sum_p(h_{k,k}(\ell)+h_{q,q}(\ell)
-2h_{p,p}(\ell))h_{k,p}(\ell)h_{p,q}(\ell).
\eeq
In the sequel we will not explicitly write down the $\ell$-dependence
of the matrix elements of $H$ and $\eta$.
Since $\eta_{k,q}=-\eta_{q,k}$, (\ref{flqm1}) describes a continuous unitary
transformation of $H$. The quantities $Tr(H^n)$ do not depend on $\ell$.
To study other properties of the flow equations (\ref{flqm2}),
let us calculate the derivative of $\sum_{k\ne q}h_{k,q}^2$.
\beqar \label{flqm3}
\frac{d}{d\ell}\sum_{k\ne q}h_{k,q}^2
&=& -\frac{d}{d\ell}\sum_{k}h_{k,k}^2
\ret
&=& -2\sum_{k,q}(h_{k,k}-h_{q,q})^2h_{k,q}^2
\ret
&=& -2\sum_{k,q}\eta_{k,q}^2
\eeqar
$\sum_{k\ne q}h_{k,q}^2$ is a monotonously decaying function of $\ell$.
Since it is bounded from below, its derivative with respect to $\ell$
vanishes if $\ell$ tends to infinity. This shows that, as already
mentioned in the introduction, $\eta_{k,q}$ vanishes in the limit
$\ell\rightarrow\infty$. Therefore, in the limit $\ell\rightarrow\infty$,
we obtain a matrix $H$ that commutes with its diagonal part $H^d$.
This means that up to degeneracies in $H^d$ the matrix $H$ has been
diagonalinot possible to solve the flow equations
(\ref{flqm2}) analytically.

In this paper we introduce an approximation to the flow equations
(\ref{flqm1}) and (\ref{etam1}). In the first step in
this approximation we assume
that a class of matrix elements may be neglected if these matrix
elements are zero initally. Suppose that the matrix elements of $H$
are divided into two classes, $C^{(1)}$ and $C^{(2)}$, such that the matrix
elements in $C^{(2)}$ vanish for $\ell=0$. The approximate
flow equations are obtained from (\ref{flqm2}) if we put
$h_{k,q}=0$ if $h_{k,q}\in C^{(2)}$. In the case of an arbitrary matrix
$H$ such an approximation is perhaps not very useful, since it is
difficult to estimate the error. But if we study the problem of e.g.
interacting electrons on lattice, we usually start with an idealized
Hamiltonian that contains a kinetic energy and a simple interaction.
Multi-particle interactions or slight modifications of the
kinetic energy are usually not expected to play an important role,
and if they do so the model has to be modified. Therefore we expect that
the physical behaviour of the model does not change too much if we
neglect matrix elements corresponding to contributions to the
Hamiltonian not included initially. Nevertheless it would be desirable
to have more general statements about the validity of this approximation.

Let us for a moment assume that initially
the off-diagonal matrix elements are small,
such that a usual perturbational treatment of $H^r$ is justified.
In this case it is possible to solve the flow equations (\ref{flqm2})
iteratively. As a first approximative solution we take
\beq \label{pertu1}
h^{(1)}_{k,q}=h_{k,q}(0)\exp(-(h_{k,k}(0)-h_{q,q}(0))^2\ell).
\eeq
The $(n+1)$-th approximation is now obtained from the $n$-th
approximation if\beq \label{pertu2}
\frac{dh^{(n+1)}_{k,q}}{d\ell}=\sum_p(h^{(n)}_{k,k}+h^{(n)}_{q,q}
-2h^{(n)}_{p,p})h^{(n)}_{k,p}h^{(n)}_{p,q}.
\eeq
In the case $n=1$ the right hand side is easily integrated and we obtain
\beqar \label{pertu3}
h^{(2)}_{k,q} &=& \sum_p\frac{h_{k,k}(0)+h_{q,q}(0)
-2h_{p,p}(0)}{(h_{k,k}(0)-h_{p,p}(0))^2+(h_{q,q}(0)-h_{p,p}(0))^2}
h_{k,p}(0)h_{p,q}(0)
\ret & &
\left[1-\exp \left(
-\left((h_{k,k}(0)-h_{p,p}(0))^2+(h_{q,q}(0)-h_{p,p}(0))^2\right)\ell
\right)\right]
\eeqar
In the limit $\ell\rightarrow\infty$, and if no degeneracy occurs
this yields
\beq \label{pertu4}
h^{(2)}_{k,k}(\infty)
=\sum_p\frac{h_{k,p}(0)h_{p,k}(0)}{h_{k,k}(0)-h_{p,p}(0)}
\eeq
which is equivalent to the result
obtained from ordinary perturbation theory.

In our approximation we neglected matrix elements $h_{k,q}\in C^{(2)}$
for which $h_{k,q}(0)=0$. Since they do not contribute to the
right hand side of (\ref{pertu4}), our approximation agrees with
a perturbational treatment up to second order in $H^r$.
But it is not necessarily restricted to the regime where
perturbation theory is valid.

The choice of $C^{(2)}$ above is completely arbitrary, and in the following
we will choose $C^{(2)}$ almost as large as possible. The
question is then whether or not the terms neglected are relevant in the
sense that they alter the solution significantly. In the second step in our
method
we then investigate
this problem by simply moving some of the elements of $C^{(2)}$ to $C^{(1)}$
and trying to analyse the new flow equations. In this way different
new contributions to the Hamiltonian will be treated and for each of
them we try to find out whether it is relevant or not.

\Section{Flow equations for the Anderson model} \label{SFAm}
\%

The Hamiltonian of the Anderson model in a normal-ordered form is given by
\cite{Anderson}
\beqar \label{HAm}
H &=& \sum_{k,\sigma}\epsilon_k:c^{\dagg}_{k,\sigma}c_{k,\sigma}:
+\sum_{\sigma}\tilde{\epsilon}_d:d^{\dagg}_{\sigma}d_{\sigma}:
+\sum_{k,\sigma}V_k(:c^{\dagg}_{k,\sigma}d_{\sigma}:
+:d^{\dagg}_{\sigma}c_{k,\sigma}:)
\ret
&+& U:d^{\dagg}_{+}d^{\dagg}_{-}d_{-}d_{+}:.
\eeqar
The first term represents the kinetic energy of the band electrons.
In addition there is a defect state. It hybridizes with the
conduction electrons. Since the phase
of $c_{k,\sigma}$ is arbitrary, we choose $V_k\ge 0$.
In a more general model, there will be
many defects in the lattice and the defect states will
be degenerate. Here we treat only the simplest case, a single,
non-degenerate defect state. Electrons in the defect state interact
due to the on-site Coulomb repulsion. This is described by the
fourth term in (\ref{HAm}). Normal order is defined by substracting
the ground state expectation values of the Hamiltonian
with $V_k=0$.
\beq
:c^{\dagg}_{k,\sigma}c_{k,\sigma}:=c^{\dagg}_{k,\sigma}c_{k,\sigma}-n_k
\eeq
\beq
:d^{\dagg}_{\sigma}d_{\sigma}:=d^{\dagg}_{\sigma}d_{\sigma}-n_d
\eeq
where
\beq
n_k = \theta(\epsilon_F-\epsilon_k)
\eeq
and
\beq
n_d = \frac{1}{2}(\theta(\epsilon_F-\epsilon_d)+
\theta(\epsilon_F-\epsilon_d-U)).
\eeq
$\tilde{\epsilon}_d$ is given by
\beq
\tilde{\epsilon}_d=\epsilon_d+n_dU
\eeq
where $\epsilon_d$ is the energy of the single occupied defect state.

\subsection{The flow equations}

We now set
\beq \label{Hr1}
H^r=\sum_{k,\sigma}V^r_k(:c^{\dagg}_{k,\sigma}d_{\sigma}:
+:d^{\dagg}_{\sigma}c_{k,\sigma}:).
\eeq
In the spirit of our approach i$V^r_k=V_k$. But we will later see that this is
too restrictive in
some cases. Still one can always think of $V^r_k=V_k$ for nearly all
values of $k$.
$\eta$ may be written as
\beqar \label{eta1}
\eta &=& [H,H^r]
\ret
&=& \sum_{k,\sigma}\eta_k(:c^{\dagg}_{k,\sigma}d_{\sigma}:
-:d^{\dagg}_{\sigma}c_{k,\sigma}:)
+\sum_{k,q,\sigma}\eta_{k,q}(:c^{\dagg}_{k,\sigma}c_{q,\sigma}:
-:c^{\dagg}_{q,\sigma}c_{k,\sigma}:)
\ret
&+& \sum_{k,\sigma}\eta^{(2)}_k(:c^{\dagg}_{k,\sigma}
d^{\dagg}_{-\sigma}d_{-\sigma}d_{\sigma}:
-:d^{\dagg}_{\sigma}d^{\dagg}_{-\sigma}d_{-\sigma}c_{k,\sigma}:)
\eeqar
where
\beq
\eta_k = (\epsilon_k-\tilde{\epsilon}_d)V^r_k
\eeq

\beq
\eta_{k,q} = \frac{1}{2}(V_kV^r_q-V_qV^r_k)
\eeq

\beq
\eta^{(2)}_k = -U\,V^r_k
\eeq
The commutator of $\eta$ and $H$ can be calculated easily.
\beqar \label{cetaH}
[\eta,H] &=& \sum_{k,\sigma}\eta_k(\tilde{\epsilon}_d-\epsilon_k)
(:c^{\dagg}_{k,\sigma}d_{\sigma}:
+:d^{\dagg}_{\sigma}c_{k,\sigma}:)
\ret
&+& \sum_{k,q,\sigma}\eta_kV_q
(:c^{\dagg}_{k,\sigma}c_{q,\sigma}:
+:c^{\dagg}_{q,\sigma}c_{k,\sigma}:)
\ret
&-& 2\sum_{k,\sigma}\eta_kV_k:d^{\dagg}_{\sigma}d_{\sigma}:
+2\sum_{k,\sigma}\eta_kV_k(n_k-n_d)
\ret
&+& U\sum_{k,\sigma}\eta_k
(:d^{\dagg}_{\sigma}d^{\dagg}_{-\sigma}d_{-\sigma}c_{k,\sigma}:
+:c^{\dagg}_{k,\sigma}d^{\dagg}_{-\sigma}d_{-\sigma}d_{\sigma}:)
\ret
&-& \sum_{k,q,\sigma}\eta_{k,q}(\epsilon_k-\epsilon_q)
(:c^{\dagg}_{k,\sigma}c_{q,\sigma}:
+:c^{\dagg}_{q,\sigma}c_{k,\sigma}:)
\ret
&+& 2\sum_{k,q,\sigma}\eta_{k,q}V_q
(:d^{\dagg}_{\sigma}c_{k,\sigma}:
+:c^{\dagg}_{k,\sigma}d_{\sigma}:)
\ret
&-& \sum_{k,\sigma}\eta^{(2)}_k(\epsilon_k-\tilde{\epsilon}_d)
(:d^{\dagg}_{\sigma}d^{\dagg}_{-\sigma}d_{-\sigma}c_{k,\sigma}:
+:c^{\dagg}_{k,\sigma}d^{\dagg}_{-\sigma}d_{-\sigma}d_{\sigma}:)
\ret
&-& \sum_{k,\s:d^{\dagg}_{\sigma}d^{\dagg}_{-\sigma}d_{-\sigma}d_{\sigma}:
\ret
&+& \sum_{k,q,\sigma}\eta^{(2)}_kV_q
(:c^{\dagg}_{k,\sigma}d^{\dagg}_{-\sigma}d_{-\sigma}c_{q,\sigma}:
+:c^{\dagg}_{q,\sigma}d^{\dagg}_{-\sigma}d_{-\sigma}c_{k,\sigma}:
-:c^{\dagg}_{k,\sigma}d^{\dagg}_{-\sigma}d_{\sigma}c_{q,-\sigma}:
\ret & &
-:c^{\dagg}_{q,\sigma}d^{\dagg}_{-\sigma}d_{\sigma}c_{k,-\sigma}:
-:c^{\dagg}_{k,\sigma}c^{\dagg}_{q,-\sigma}d_{-\sigma}d_{\sigma}:
-:d^{\dagg}_{\sigma}d^{\dagg}_{-\sigma}c_{q,-\sigma}c_{k,\sigma}:)
\ret
&+& 2\sum_{k,\sigma}\eta^{(2)}_kV_k
(n_k-n_d):d^{\dagg}_{-\sigma}d_{-\sigma}:
\ret
&+& U\sum_{k,\sigma}\eta^{(2)}_k(1-2n_d)
(:d^{\dagg}_{\sigma}d^{\dagg}_{-\sigma}d_{-\sigma}c_{k,\sigma}:
+:c^{\dagg}_{k,\sigma}d^{\dagg}_{-\sigma}d_{-\sigma}d_{\sigma}:)
\ret
&+& U\sum_{k,\sigma}\eta^{(2)}_kn_d(1-n_d)
(:d^{\dagg}_{\sigma}c_{k,\sigma}:+:c^{\dagg}_{k,\sigma}d_{\sigma}:)
\eeqar
To obtain the flow equations for the matrix elements of $H$ one has to
compare the different terms on the right hand side of (\ref{cetaH}) with
$H$. For instance, the second term in (\ref{cetaH})
is the only term that contributes to the derivative of $\epsilon_k$
with respect to $\ell$,
\beq \label{flk}
\frac{d\epsilon_k}{d\ell}=2\eta_kV_k=2(\epsilon_k-\tilde{\epsilon}_d)
V_kV_k^r.
\eeq
All the terms containing $:d^{\dagg}_{\sigma}d_{\sigma}:$ contribute
to the derivative of $\tilde{\epsilon}_d$ with respect to $\ell$.
But, since $\tilde{\epsilon}_d=\epsilon_d+n_dU$ and $n_d$ eventually
changes with $\ell$, there is an extra contribution $U\frac{dn_d}{d\ell}$.
Therefore
\beqar \label{fld}
\frac{d\tilde{\epsilon}_d}{d\ell} &=& -2\sum_k\eta_kV_k
+2\sum_k\eta_k^{(2)}V_k(n_k-n_d)+U\frac{dn_d}{d\ell}
\ret
&=& 2\sum_k(\tilde{\epsilon}_d-\epsilon_k)V_kV_k^r
+2U\sum_kV_kV_k^r(n_d-n_k)+U\frac{dn_d}{d\ell}.
\eeqar
Simila\beqar \label{flV}
\frac{dV_k}{d\ell} &=&-\eta_k(\epsilon_k-\tilde{\epsilon}_d)
+2\sum_p\eta_{k,p}V_p+Un_d(1-n_d)\eta_k^{(2)}
\ret
&=& -V_k^r(\epsilon_k-\tilde{\epsilon}_d)^2
+V_k\sum_pV_pV_p^r
-V_k^r\sum_pV_pV_p
-U^2n_d(1-n_d)V_k^r
\eeqar
and
\beq \label{flU}
\frac{dU}{d\ell}=-4\sum_k\eta_k^{(2)}V_k=4U\sum_kV_kV_k^r.
\eeq
We will now analyse these equations.

\subsection{The symmetric case}

If
$\epsilon_{\pi-k}=-\epsilon_k$,
$V_{\pi-k}=V_k$,
$\tilde{\epsilon}_d=0$,
$\epsilon_F=0$,
$n_d=\frac{1}{2}$ the Hamiltonian is invariant under particle hole
transformations
\beq
c^{\dagg}_{k,\sigma}\,\rightarrow\,-c_{\pi-k,\sigma},\quad
c_{k,\sigma}\,\rightarrow\,-c^{\dagg}_{\pi-k,\sigma},\quad
d^{\dagg}_{\sigma}\,\rightarrow\,d_{\sigma},\quad
d_{\sigma}\,\rightarrow\,d^{\dagg}_{\sigma}.
\eeq
We choose $V_k=V_k^r$ and obtain
\beqar \label{flqsymm}
\frac{d\epsilon_k}{d\ell} &=& 2\epsilon_kV_k^2
\ret
\frac{dV_k}{d\ell} &=& -(\epsilon_k^2+\frac{1}{4}U^2)V_k
\ret
\frac{dU}{d\ell} &=& 4U\sum_kV_k^2
\eeqar
Due to the particle-hole symmetry $\tilde{\epsilon}_d$ remains zero.
The equations (\ref{flqsymm})
show that $V_k$ tends to zero for all $k$, whereas
$\abs{U}$ and $\abs{\epsilon_k}$ increase.
For the density of states at the Fermi level $\epsilon_F=0$,
\beq
\rho(0,l)=\frac{1}{(2\pi)^d}\int d^dk\delta(\epsilon_k),
\eeq
we obtain
\beq
\frac{d\rho(0,l)}{d\ell}=-2V^2\exp(-\frac{1}{2}\int_0^{\ell}U^2d\ell)
\rho(0,l).
\eeq
Here $V$ is the initial value of $V_k$ for those $k$ with
$\epsilon_k=0$.
The density of states in the middle of the band decreases, but it
does not vanish unless $U=0$.

The case $U=0$ must be treated seperately. In this case we obtain
\beq
\epsilon_k(\infty)=\mbox{sign}(\epsilon_k(0)) \sqrt{\epsilon_k^2(0)+2V_k^2(0)}
\eeq
and $V_k(\infty)=0$ if $\epsilon_k(\infty)\neq 0$.
If $V_k\neq0$ for such $k$ with $\epsilon_k$ near the Fermi energy,
we obtain a finite gap in the energy band with some states in the
middle of the gap. A gap does not show up for a single impurity
in the thermodynamic limit, since $V_k \sim N_s^{-\frac{1}{2}}$
where $N_s$ is the number of lattice sites. On the other hand, a system
with a small but finite density of defects behaves like a system
with a single defect in a finite volume, the volume per defect.
The gap would be of the order of the density of defects.
This result is presumably an artefact of
our approximation, since for $U=0$ the equations (\ref{flqsymm})
decouple. If $\epsilon_k$ is near the Fermi energy,
$V_k$ tends to zero quite slowly and therefore other matrix elements
neglected so far are important. Furthermore, the result for $U=0$ is
not stable with respect to small changes of the Hamiltonian.
If one introduces e.g. a non-vanishing interaction $U$, the system
has no gap as expected.

\subsection{The asymmetric case}

In general the system has no particle hole symmetry. The flow equations
are given by
\beq \label{flak}
\frac{d\epsilon_k}{d\ell}=2(\epsilon_k-\epsilon_d-n_dU)
V_kV_k^r
\eeq

\beq \label{flad}
\frac{d\epsilon_d}{d\ell}=
2\sum_k(\epsilon_d-\epsilon_k-n_kU)V_kV_k^r
\eeq

\beq \label{flaVk}
\frac{dV_k}{d\ell}=
-(\epsilon_k-\epsilon_d-n_dU)^2V_k^r
+V_k\sum_pV_pV_p^r
-V_k^r\sum_pV_pV_p
-U^2n_d(1-n_d)V_k^r
\eeq

\beq \label{flaU}
\frac{dU}{d\ell}=4U\sum_kV_kV_k^r
\eeq
Now $V_k$ does not necessarily vanish for all $k$ and we have to choose
\beq
V_k^r=g_k^rV_k
\eeq
where
\beqar
g_k^r &=& 0\quad\mbox{if}\quad(\epsilon_k-\epsilon_d-n_dU)^2+n_d(1-\rightarrow
0 \quad\mbox{for}\quad \ell\rightarrow\infty
\ret
g_k^r &=& 1\quad\mbox{otherwise.}
\eeqar
With this choice of $V_k^r$ we ensure that the right hand side of
(\ref{flaU}) vanishes for $\ell\rightarrow\infty$. Otherwise $U$
becomes infinite for $\ell\rightarrow\infty$. Such a divergence is
typical for approximate flow equations. A similar divergence occured
also in the paper of Wegner \cite{Wegner}. It is clear that the original
equation (\ref{flq1}) contains no such divergences since it describes a
continous unitary transformation of the Hamiltonian. But it is always
possible to choose $H^r$ such that it vanishes for $\ell\rightarrow\infty$,
and with such a choice no divergences occur.

We have to distiguish four possible cases:
\begin{enumerate}
\item $n_d=\frac{1}{2}$ for $\ell\rightarrow\infty$. This case is similar
to the symmetric case discussed above. All $V_k$ vanish, $\abs{U}$
increases and the energies $\epsilon_k$ and $\epsilon_d$ are renormalized.
The defect state is occupied with a single electron, it has a magnetic
moment. This case is therefore called the local moment fixed point.
\item $n_d=0$ or $n_d=1$ for $\ell\rightarrow\infty$ and
$g_k^r=0$ for some $k$. In this case the corresponding $V_k$
do not vanish. There is a non-vanishing coupling of the defect state
to the states in the band for which $\epsilon_k=\epsilon_d+n_dU$. Since
the corresponding $V^r_k$ is zero, only the second term in
(\ref{flaVk}) contributes and $\abs{V_k}$ increases. This case
is called the strong coupling fixed point.
\item $n_d=0$ or $n_d=1$ for $\ell\rightarrow\infty$ and
$g_k^r\ne 0$ for all $k$. In this case $\epsilon_d$ and $\epsilon_d+U$
are both above (for $n_d=0$) or below (for $n_d=1$) the band. The defect
state is decoupled from the band and the electrons in the band behave like
free elect\item The non-interacting case with
$n_d=0$, $n_d=\frac{1}{2}$, or $n_d=1$. In this case $U=0$
and $V_k$ vanishes if $\epsilon_k\ne\tilde{\epsilon}_d$
\end{enumerate}

The above mentioned fixed points are not fixed points in the sense that
all the parameter of the model are fixed. They describe merely classes
of parameters which show a similar physical behaviour.
The main question is into which of the four classes the system falls,
depending on the initial Hamiltonian. The last case is simple since it
describes an instable fixed point of the flow equations. Only if $U=0$
initially it remains zero for all $\ell$.
The other fixed points are more interesting,
we will discuss them in detail.

The first case is the local moment fixed point, where $n_d=\frac{1}{2}$,
e.g. $\epsilon_d<\epsilon_F$ and $\epsilon_d+U>\epsilon_F$
(the case $\epsilon_d>\epsilon_F$ and $\epsilon_d+U<\epsilon_F$ is similar).
We shall assume that $U$ is not too small initially. Now the last term
in (\ref{flaVk}) is the most important term.
Using $U\ge U(0)$ it already yields the estimate
\beq \label{ubV}
V_k\le V_k(0)\exp(-\frac{1}{4}U^2(0)\ell).
\eeq
Inserting this in (\ref{flaU}) we obtain
\beq
U \le U(0)\exp \left(8\sum_k \frac{V_k^2(0)}{U^2(0)}
[1-\exp (-\frac{1}{2}U^2(0)\ell)]\right).
\eeq
In a typical situation, $\sum_k \frac{V_k^2(0)}{U^2(0)}$ is of order 1
or smaller. Now, if
$\abs{\epsilon_k-\epsilon_d-\frac{1}{2}U}\le c_k$
we obtain
\beq
\abs{\epsilon_k-\epsilon_k(0)}\le \frac{8c_kV_k^2(0)}{U^2(0)}
[1-\exp (-\frac{1}{4}U^2(0)\ell)]
\eeq
This shows that $\epsilon_k$ does not change very much. Therefore if
we start with a symmetric conduction band, it remains essentially
symmetric and $\sum_k \epsilon_k V_k^2 \approx 0$.
Consequently the right hand side of (\ref{flad})$\epsilon_d$ remains below
$\epsilon_F$. Only if the conduction
band is strongly asymmetric, the system may behave differently.
The derivative of $\epsilon_d+U$ is approximately given by
$\sum_k(\epsilon_d+(1-n_k)U)V_k^2$. Thus if $\epsilon_d$ becomes of
the order $-(1-\rho/2)U$, where $\rho$ is the density of electrons,
$\epsilon_d+U$ decreases and the
system may change to a state where $n_d=1$. Otherwise it remains in a
state with $n_d=\frac{1}{2}$. The precise values of the parameters
where the transition occurs cannot be determined using the rough
estimates above.

The situation becomes more complicated if initially $n_d=0$,
i.e. $\epsilon_d>\epsilon_F$ and $\epsilon_d+U>\epsilon_F$.
If $\epsilon_d$ is only slightly above $\epsilon_F$, the right hand side of
(\ref{flad}) will still be negative so that $\epsilon_d$ decreases
below $\epsilon_F$. Then $n_d=\frac{1}{2}$ and the system rests in the
local moment fixed point. On the other hand, if $\epsilon_d>U$ and
if we start with a symmetric conduction band, the right hand side of
(\ref{flad}) is positive and $\epsilon_d$ increases. Somewhere between
these two possibilities there must be a transition where the system,
depending on the inital values of the parameters, switches from the
local moment fixed point to a fixed point with $n_d=0$. The question,
whether this fixed point is the free electron fixed point or
the strong coupling fixed point, cannot be decided easily.

Numerical solutions of the equations are presented below, they confirm
this qualitative discussion.

\subsection{Numerical results} \label{NRes}

We have implemented an adaptive stepsize fifth order Runge--Kutta algorithm to
solve the flow equations (\ref{flak}-\ref{flaU}) numerically. All calperformed
for a symmetric band\linebreak$\epsilon_k=-1+2|k|$, $k\in[-1,1]$,
$\epsilon_F=0$, one impurity site and $N_s=50$ sites in the conduction band.
The choice of the discrete $N_s$~values of~$k$ in the conduction band does not
affect
the results very much. We took them equidistant in the interval $[-1,1]$.
In the thermodynamic limit these parameters correspond to a 2\%~density of
impurities and a
constant density of states in the conduction band. The hybridization
energies scale like $V_k=V(0)/\sqrt{N_s}$ in this limit. We have set $V(0)=2$.

Figures 1 to 4 show  numerical solutions of the flow equations for a starting
value
$U(0)=5$. Corresponding to the initial value of~$\epsilon_d$, one finds e.g.
the
local moment fixed point (figures~1 and~2) or the free electron fixed point
(figures~3 and~4). In either case the off--diagonal elements~$V_k$ vanish in
the
limit $l\rightarrow\infty$ as intended by the continuous unitary
transformations.
One notices in figure~2 that this convergence of the~$V_k$ gets much faster
once $\epsilon_d$ is below the Fermi level. This is due to the last term
in (\ref{flaVk}).

In the free electron case the convergence of some~$V_k$ is much
slower since we are coming close to the crossover to the strong coupling fixed
point. For the same reason $U$ gets very large. But this could only be noticed
if the impurity site were twice occupied, which is a very high lying excitation
for
this fixed point anyway that we cannot hope to describe by the model.
The case shown in figures~3 and~4 lies near the crossover to the
strong coupling fixed point. If $\epsilon_d(0)$ is larger, the
convergence is faster and the final value of $U$ is smaller. If
$\epsilon_d(0)$ is smaller, $\epsilon_d(\ell)$ intersects
the curve of the first band energy and $\epsilon_d(\infty)$
lies in the band.
 Finally one can wonder about the attraction regions belonging to the different
fixed points. There are four parameters in our model: The bandwidth of
the conduction band, the initial values of $\epsilon_d$, $U$ and of the average
hybridization~$V$. Obviously only three of these
parameters can be independent since the energy scale is arbitrary. In our case
the bandwidth is fixed. Of the remaining three parameters only two are really
independent because the flow equations connect unitarily equivalent
Hamiltonians
with different values of these parameters (the bandwidth hardly changes as
mentioned
before for this small density of impurities). Therefore we can restrict
ourselves to
investigating the attraction regions for a fixed value of~$V(0)$, here
$V(0)=2$.

In figure 5 the approximate boundaries of the parameter sets belonging to the
local moment or the free electron behaviour are drawn in the
$\left(\epsilon_d(0)\,,U(0)\right)$--plane. In between lies the region
of the strong coupling fixed point. Since our numerical algorithm converges
very
slowly in this region (compare figure~4), figure~5 shall mainly give a
qualitative impression of this region.

\Section{Relevant and irrelevant operators} \label{SRio}

In the commutator $[\eta,H]$, many terms have been
neglected and it is not clear
a priori whether or not these terms are relevant for the physical behaviour
of the model. To investigate this question, we add contributions
of this type to the Hamiltonian and write down the flow equations
for the new Hamiltonian. A detailed derivation of the new flow
equations is given in the appendix.

A simple contribution in $[\eta,H]$ that has been neglected
is of the form
$\sum_{k,q,\sigma}V_{k,q}:c^{\dagg}_{k,\sigma}c_{q,\sigma}:$.It will turn out
that such a contribution is irrelevant for the model.
To see this we have to investigate the flow equations for a Hamiltonian
including such a term, i.e.
\beqar
H &=& \sum_{k,\sigma}\epsilon_k:c^{\dagg}_{k,\sigma}c_{k,\sigma}:
+\sum_{\sigma}\tilde{\epsilon}_d:d^{\dagg}_{\sigma}d_{\sigma}:
+\sum_{k,\sigma}V_k(:c^{\dagg}_{k,\sigma}d_{\sigma}:
+:d^{\dagg}_{\sigma}c_{k,\sigma}:)
\ret
&+& U:d^{\dagg}_{+}d^{\dagg}_{-}d_{-}d_{+}:
+\sum_{k,q,\sigma}V_{k,q}:c^{\dagg}_{k,\sigma}c_{q,\sigma}:.
\eeqar
The flow equations are obtained from (\ref{Afli1k}-\ref{Afli1U})
\beq \label{fli1k}
\frac{d\epsilon_k}{d\ell} = 2(\epsilon_k-\tilde{\epsilon}_d)V_kV^r_k
+2\sum_q(\epsilon_k-\epsilon_q)V_{k,q}^2
+2\sum_qV_{k,q}(2V^r_qV_k-V_qV_k-V_qV^r_k)
\eeq
\beqar \label{fli1d}
\frac{d\tilde{\epsilon}_d}{d\ell}
&=& 2\sum_k(\tilde{\epsilon}_d-\epsilon_k)V_kV_k^r
+2U\sum_kV_kV_k^r(n_d-n_k)
\ret & &
+\frac{1}{2}\sum_{k,q}V_{k,q}(V_q-V^r_q)(V_k-V^r_k)
+U\frac{dn_d}{d\ell}
\eeqar
\beqar \label{fli1V}
\frac{dV_k}{d\ell}
&=& -V^r_k((\epsilon_k-\tilde{\epsilon}_d)^2+n_d(1-n_d)U^2)
+\sum_p(\epsilon_k-\epsilon_p)V_{k,p}V_p
+\sum_p(\tilde{\epsilon}_d-\epsilon_p)V_{k,p}V^r_p
\ret & &
+\sum_p(V_kV^r_p-V_pV^r_k)V_p
+\sum_{q,p}V_{k,p}V_{p,q}(V_q-V^r_q)
\eeqar
\beqar \label{fli1Vkq}
\frac{dV_{k,q}}{d\ell}
&=& -(\epsilon_k-\epsilon_q)^2V_{k,q}
+(\epsilon_k-\tilde{\epsilon}_d)V^r_kV_q
+(\epsilon_q-\tilde{\epsilon}_d)V_kV^r_q
+\sum_p(\epsilon_k+\epsilon_q-2\epsilon_p)V_{k,p}V_{p,q}
\ret & &
-\frac{1}{2}(V_kV^r_q-V_qV^r_k)(\epsilon_k-\epsilon_q)
+\sum_p[(V_kV^r_p-V_pV^r_k)V_{p,q}+(V_qV^r_p-V_pV^r_q)V_{p,k}]
\ret & &
+\sum_p(V^r_p-V_p)[V_{k,p}V_q+V_{q,p}V_k]
\eeqar
\beq \label{fli1U}
\frac{dU}{d\ell}=-4\sum_k\eta_k^{(2)}V_k=4U\sum_kV_k^2
\eeq
First, one observes
$V_{k,q}(\infty)=0$ if $(\epsilon_k-\epsilon_q)\neq0$. $V_{k,q}$does not vanish
for $\ell\rightarrow\infty$ if
$\epsilon_k=\epsilon_q$. But this does not cause any problems, since
nevertheless the second term in (\ref{fli1k}) tends to zero for
$\ell\rightarrow\infty$.
There are no diverging matrix elements in this limit.
The equation for $U$ is the same as before and $\abs{U}$ increases.
In the equation for $V_k$ additional contributions occur, but
they vanish for $\ell\rightarrow\infty$ so that $V_k$ vanishes
if $n_d=\frac{1}{2}$ or $\epsilon_k\ne\tilde{\epsilon}_d$.
Finally, the additional contribution
$\sum_{k,q,\sigma}V_{k,q}:c^{\dagg}_{k,\sigma}c_{q,\sigma}:$
to the Hamiltonian does not change the behaviour of the system.
$\epsilon_k$, $\tilde{\epsilon}_d$ and $U$ are "`renormalized"' somewhat,
but the behaviour of the system remains the same.
Only in the non-interacting case is
$\sum_{k,q,\sigma}V_{k,q}:c^{\dagg}_{k,\sigma}c_{q,\sigma}:$
a relevant contribution to the Hamiltonian.

The next contribution in $[\eta,H]$ that was neglected is a term of the form
$\sum_{k,\sigma}V_k^{(2)}
(:c^{\dagg}_{k,\sigma}d^{\dagg}_{-\sigma}d_{-\sigma}d_{\sigma}:
+:d^{\dagg}_{\sigma}d^{\dagg}_{-\sigma}d_{-\sigma}c_{k,\sigma}:)$.
Such a contribution only occurs if $U\ne0$. But if $U\ne 0$ initially
$U$ always increases. Therefore we should expect that matrix
elements between states with $n_d=\frac{1}{2}$ and $n_d=1$ are not
important. This means that this contribution is likely to be irrelevant
for all fixed points. To see this we add such a term to $H$ and $H^r$.
The flow equations are then
\beq \label{fli2k}
\frac{d\epsilon_k}{d\ell}
=2(\epsilon_k-\tilde{\epsilon}_d)(V^r_kV_k+n_d(1-n_d)V^{(2)2}_k)
-2n_d(1-n_d)U\,V^{(2)}_k(V_k+V^r_k)
\eeq
\beqar \label{fli2d}
\frac{d\tilde{\epsilon}_d}{d\ell}
&=& -2\sum_k(\epsilon_k-\tilde{\epsilon}_d+U(n_k-n_d))
(V^r_kV_k+n_d(1-n_d)V^{(2)2}_k)
\r+2\sum_k((\epsilon_k-\tilde{\epsilon}_d)(n_k-n_d)+n_d(1-n_d)U)
(V^r_k+V_k)V_k^{(2)}
\ret & &
-2(1-2n_d)U\sum_k(n_k-n_d)V_kV^{(2)}_k
+U\frac{dn_d}{d\ell}
\eeqar
\beqar \label{fli2Vk}
\frac{dV_k}{d\ell}
&=&
-((\epsilon_k-\tilde{\epsilon}_d)^2+n_d(1-n_d)U^2)V^r_k
+2n_d(1-n_d)(\epsilon_k-\tilde{\epsilon}_d)U\,V^{(2)}_k
\ret & &
+\sum_p\left((V_kV^r_p-V_pV^r_k)V_p
+n_d(1-n_d)[(V_k-V^r_k)V^{(2)}_p-(V_p-V^r_p)V^{(2)}_k]V_p^{(2)}\right)
\eeqar
\beqar \label{fli2U}
\frac{dU}{d\ell}
&=&4U\,\sum_kV_k^rV_k
-4\sum_k((\epsilon_k-\tilde{\epsilon}_d)-(1-2n_d)U)V^{(2)}_k(V_k^r+V_k)
\ret & &
+4\sum_k[(1-3n_d(1-n_d))U-(1-2n_d)(\epsilon_k-\tilde{\epsilon}_d)]
V^{(2)2}_k
\eeqar
\beqar\label{fli2V2}
\frac{dV_k^{(2)}}{d\ell}
&=& -[n_d(1-n_d)U^2+(\epsilon_k-\tilde{\epsilon}_d-(1-2n_d)U)^2]V^{(2)}_k
\ret & &
+[2U(\epsilon_k-\tilde{\epsilon}_d)-(1-2n_d)U^2]V_k^r+
\ret & &
+\sum_q((V_k-V^r_k)V^{(2)}_q-(V_q-V^r_q)V^{(2)}_k)(V_q+(1-2n_d)V_q^{(2)})
\ret & &
+\sum_q(V_kV^r_q-V_qV^r_k)V_q^{(2)}
\eeqar
Although these equations look complicated, they are easy to analyse.
Let us first look at (\ref{fli2U}). $U$ increases as long as
$V_k^r$ and $V_k^{(2)}$ do not vanish. On the other hand, the first
term in (\ref{fli2V2}) guarantees that $V_k^{(2)}$ tends to zero
in the limit $\ell\rightarrow\infty$. For sufficiently large $U$
the dominant contribution is $\sim -U^2V^{(2)}_k$ so that
$V^{(2)}_k$ vanishes rapidly. This shows that for $n_d=\frac{1}{2}$
the behaviour of $V^{(2)}_k$ is similar to the behaviour of
$V_k$. In (\ref{fli2k}), (\ref{fli2d}) and (\ref{fli2U}) the additional
terms containing $V_kV^{(2)}_k$ and $V^{(2)2}_k$ are similar to
the original contributions containing $V_k^2$. This shows that
the only effect of the additional contribution
$\sum_{k,\sigma}V_k^{(2)}
(:c^{\dagg}_{k,\sigma}d^{\dagg}_{-\sigma}d_{-\si+:d^{\dagg}_{\sigma}d^{\dagg}_{-\sigma}d_{-\sigma}c_{k,\sigma}:)$
to the Hamiltonian is a renormalization of $\epsilon_k$,
$\tilde{\epsilon}_d$, and $U$, which also occurs if one changes
e.g. the initial values of $V_k$.
If $n_d=0$ or $n_d=1$ we saw that $V_k$ does not necessarily vanish
for $\ell\rightarrow\infty$. In contrast $V^{(2)}_k$ tends to zero
in this case too. Furthermore the contributions on the right hand side
of (\ref{fli2k}) and (\ref{fli2Vk}) containing $V_k^{(2)}$ vanish.
Consequently, the behaviour of $\epsilon_k$ and $V_k$ does not
change significantly. Only $\tilde{\epsilon}_d$ and $U$ are
renormalized.

In the non-interacting case ($U=0$) the inhomogeneity in (\ref{fli2V2})
vanishes. Consequently, $V_k^{(2)}=0$ if initially $V_k^{(2)}(0)=0$.
This shows that the additional contribution
$\sum_{k,\sigma}V_k^{(2)}
(:c^{\dagg}_{k,\sigma}d^{\dagg}_{-\sigma}d_{-\sigma}d_{\sigma}:
+:d^{\dagg}_{\sigma}d^{\dagg}_{-\sigma}d_{-\sigma}c_{k,\sigma}:)$
to the Hamiltonian is irrelevant for the fixed points described
above.

Another contribution on the right hand side of (\ref{cetaH}) contains
$(:c^{\dagg}_{k,\sigma}c^{\dagg}_{q,-\sigma}d_{-\sigma}d_{\sigma}:
+:d^{\dagg}_{\sigma}d^{\dagg}_{-\sigma}c_{q,-\sigma}c_{k,\sigma}:)$.
This contribution may be analysed similarly and it turns out that
it is irrelevant. This should have been expected since, as in the
previous case, such a term contains matrix elements between two states with
different $n_d$. Since $U$ becomes large, these matrix elements
tend to zero rapidly.

The last contribution in (\ref{cetaH}) not taken into account until now
is more important. It is of the form
\beq \label{Vkq2}
\sum_{k,q,\sigma}V_{k,q}^{(2)}
(:c^{\dagg}_{k,\sigma}d^{\dagg}_{-\sigma}d_{-\sigma}c_{q,\sigma}:
-:c^{\dagg}_{k,\sigma}d^{\dagg}_{-\sigma}d_{\sigma\eeq
and may be written as a linear combination
of a spin-spin interaction and a density-density interaction.
A contribution of this form was first obtained by Schrieffer and Wolff
\cite{SW}. They introduced a unitary transformation in order to
eliminate the matrix elements $V_k$. As a result of this transformation
one obtains a complicated Hamiltonian, which reduces to the original
one with $V_k=0$ and an additional contribution of the form (\ref{Vkq2})
if $\abs{V_k}$ is small.
One should expect that these terms are important in our approach
as well. To see this, we add such a term to $H$ and
$H^r$ in (\ref{HAm}) and (\ref{Hr1}).
The flow equations are obtained as before.
\beq \label{fli4k}
\frac{d\epsilon_k}{d\ell}=2(\epsilon_k-\tilde{\epsilon}_d)V^r_kV_k
-\sum_qV_{k,q}^{(2)}(V^r_q-V_q)V_k(n_q-n_d)
\eeq

\beqar \label{fli4d}
\frac{d\tilde{\epsilon}_d}{d\ell}
&=&
-2\sum_k(\epsilon_k-\tilde{\epsilon}_d)V^r_kV_k
+2\sum_{k,q}V_{k,q}^{(2)}(V^r_q-V_q)V_k(n_k+n_q-2n_d)
\ret & &
-2U\sum_kV_kV^r_k(n_k-n_d)+U\frac{dn_d}{d\ell}
\ret & &
-\sum_{k,q}(V_kV^r_q-V_qV^r_k)V_{k,q}^{(2)}(n_k-n_q)
-2(1-2n_d)\sum_{k,q,\sigma}(\epsilon_k-\epsilon_q)(n_k-n_q)V_{k,q}^{(2)2}
\eeqar

\beqar \label{fli4Vk}
\frac{dV_k}{d\ell}
&=&-[(\epsilon_k-\tilde{\epsilon}_d)^2+n_d(1-n_d)U^2]V^r_k
\ret & &
+\sum_q(\epsilon_k-\epsilon_q)V_{k,q}^{(2)}V^r_q(n_q-n_d)
+\sum_q(\tilde{\epsilon}_d-\epsilon_q)V_{k,q}^{(2)}V_q(n_q-n_d)
\ret & &
+\sum_q(V_kV^r_q-V_qV^r_k)V_q+Un_d(1-n_d)\sum_qV_{k,q}^{(2)}(3V^r_q-V_q)
\ret & &
+2\sum_{q,p}V_{k,q}^{(2)}V_{p,q}^{(2)}(V_p-V_p^r)
[2n_pn_q(1-n_d)+n_p+n_d(1-2n_d)]
\eeqar

\beq \label{fli4U}
\frac{dU}{d\ell}=4U\sum_kV_kV_k^r
-\sum_{k,q}(\epsilon_k-\epsilon_q)(n_k-n_q)V_{k,q}^{(2)2}
+4\sum_qV_{k,q}^{(2)}(V_q-V_q^r)V_k
\eeq

\beqar \label{fli4V2}
\frac{dV_{k,q}^{(2)}}{d\ell}
&=&-(\e-U(V_k^rV_q+V_q^rV_k)
-2\sum_pV_p(V_p-V_p^r)V_{k,q}^{(2)}
\ret & &
+\sum_p((2V_kV_p^r-V_pV_k^r-V_pV_k)V_{p,q}^{(2)}
+(2V_qV_p^r-V_pV_q^r-V_pV_q)V_{p,k}^{(2)})
\ret & &
+2(1-2n_d)\sum_p(\epsilon_k+\epsilon_q-2\epsilon_p)V_{k,p}^{(2)}V_{p,q}^{(2)}
\eeqar
To analyse these equations let us first consider the case $n_d=\frac{1}{2}$.
In this case we have $V_k^r=V_k$ for all $k$ and consequently only
the first two terms on the right hand side of (\ref{fli4V2}) do not
vanish. The second term is the inhomogeneity, it vanishes in the
limit $\ell\rightarrow\infty$, but it yields a non-vanishing contribution
to $V_{k,q}^{(2)}$ for finite $\ell$. Due to the first term, this contribution
will tend to zero for $\ell\rightarrow\infty$ if $\epsilon_k \ne \epsilon_q$.
But if $\epsilon_k=\epsilon_q$, $V_{k,q}^{(2)}$ remains finite. It represents
an antiferromagnetic interaction of the local moment in the defect state with
electrons in the band. This antiferromagnetic exchange coupling is well
known from the Kondo problem.
In (\ref{fli4U}) the third term vanishes. The second
term is positive and tends to zero if $\ell$ goes to infinity.
Consequently the resulting $U(\infty)$ will be somewhat larger than before.
As before, $V_k$ vanishes rapidly due to the first term, $\epsilon_k$
and $\tilde{\epsilon}_d$ are renormalized somewhat. This shows that
in the case of the local moment fixed point, the additional contribution
(\ref{Vkq2}) is marginal. It is important for the physical behaviour
of the system, but the other parameters of the model are not changed in
such a way that the system behaves completely different.
We already mentioned that Schrieffer and Wolff obtained a term similar to
(\ref{Vkq2}). To be able to compare our result with the result
in \cite{SW}, we restrict ourselves to the symmetri$k$-vectors near the Fermi
surface. Then (\ref{fli4V2}) yields
\beq \label{fli4V2a}
\frac{dV^{(2)}_{k_F,k_F}}{d\ell}=-2UV^2_{k_F}.
\eeq
If $V^2_k$ are small we can neglect contributions of higher order
in $V_k$ as in \cite{SW}. Therefore we have $U\approx U(0)$ and
consequently
$V_{k_F}\approx V_{k_F}(0)\exp(-\frac{1}{4}U^2(0)\ell)$.
Inserting this in (\ref{fli4V2a}) we obtain
\beq
V^{(2)}_{k_F,k_F}(\infty)=-4\frac{V^2_{k_F}(0)}{U}
\eeq
which is exactly the result in \cite{SW}. In the asymmetric case
the analysis is more difficult but we expect that our result
differs from their result in \cite{SW}. Especially if $\epsilon_d$
lies in the conduction band, the Schrieffer-Wolff transformation is not
well defined in contrary to our transformation. We would like to mention
that the Schrieffer-Wolff transformation leads to other contributions
to $H$ similar to the irrelevant trems in our case. These additional
contributions were neglected in \cite{SW}.

In the strong coupling fixed point, we have
$n_d=0$ or $n_d=1$. Furthermore $V_k^r=0$ for some values of $k$.
Consequently, due to the third and fourth term on the right hand side
of (\ref{fli4V2}), $V_{k,q}^{(2)}\rightarrow 0$ if $\ell\rightarrow\infty$
for all $k$ and $q$. In this case the other parameters of the model are
changed somewhat, but the contribution (\ref{Vkq2}) vanishes.
This is in contrast to the free electron fixed point, where
$n_d=0$ or $n_d=1$ but $V_k^r=V_k$ for all $k$. Here $V_{k,q}^{(2)}$
does not vanish for some values of $k$ and $q$. But there
is no magnetic moment in the defect state and therefore an antiferromagnetic
interaction plays no role. To summarize, the contribution
(\ref{Vkq2}) is irrelevant in the strong coupling fixed point
and in the free electron fixed point.

\Section{D
In the two preceeding sections we calculated  and analysed
flow equations for the Anderson model. Especially we obtained
several fixed points. These fixed points and the corresponding
relevant, marginal or irrelevant operators may be compared with
results obtained using renormalization methods by Krishnamurty,
Wilkins and Wilson \cite{KWW}. But one has to be careful
since the idea of a fixed point differs in both cases.
In a renormalization group treatment a fixed point is a single
point in the parameter space, whereas in our case a fixed point
corresponds to a class of points in the parameter space. It will turn out that
the fixed points in \cite{KWW} are prototypes in the classes
we obtain. Furthermore we should mention that in our
notation, $n_d$ is a factor of $2$ smaller than in \cite{KWW}.

\begin{enumerate}

\item The free electron fixed point.

In this case we obtain either
$\epsilon_d, \epsilon_d+U>\epsilon_k$ for all $k$, $n_d=0$,
or $\epsilon_d, \epsilon_d+U<\epsilon_k$ for all $k$, $n_d=1$.
Furthermore $V_k=0$ for all $k$. This fixed point is stable,
all the operators we discussed in section \ref{SRio} are irrelevant.
The case $n_d=0$ corresponds to the frozen-impurity fixed point
in \cite{KWW}, which is also stable and has only irrelevant operators.
In \cite{KWW} this fixed point
in characterized by $\epsilon_d\rightarrow\infty$,
$U=0$ and $V_k=0$. The physical behaviour of such a system is
a prototype of the class of final parameters we called
free electron fixed point.

\item The strong coupling fixed point.

In this case $n_d=0$ or $n_d=1$, but $\epsilon_d$ or $\epsilon_d+U$
lies in the conduction band. The hybridization $V_k$ vanishes unless
$\tilde{\epsilon}_d=\epsilon_k$, for these values of $k$
the final vAll the other operators are irrelevant in this case. It
corresponds to the strong coupling fixed point in \cite{KWW}
where some of the $V_k$ tend to infinity for fixed $\epsilon_d$
and~$U$.

\item The local moment fixed point.

In this case we have $n_d=\frac{1}{2}$. The defect state is occupied
with a single electron, representing the local moment. $V_k=0$ for
all $k$ in this case. The symmetric case falls into this class too.
We found a marginal operator which describes an antiferromagnetic
interaction of the local moment in the defect state with the
electrons in the conduction band and a density-density interaction
of the electron in the defect state with the electrons in the band.
Due to this interaction
a singlet is formed. The singlet formation takes place with electrons
in the band for which $\epsilon_k=\tilde{\epsilon}_d$. The energy gain
due to the singlet formation gives the Kondo temperature. If the
temperature is above the Kondo temperature, triplet states are
occupied. This is the usual explanation of the Kondo effect.
The local moment fixed point was found in \cite{KWW} too,
the antiferromagnetic interaction is marginal.

\item The non-interacting fixed point.

It is given by $U=0$. It is unstable with respect to the
electron-electron interaction in the defect state. Other
relevant operators are the additional hybridization of
band electrons and the antiferromagnetic coupling between
the defect state and the band electrons. The special case
where $\epsilon_d=0$ and $V_k=0$ is called free-orbital
fixed point in \cite{KWW}. It is unstable with respect
to the operators mentioned above.
\end{enumerate}

In \cite{KWW} another fixed point is mentioned, the so
called valence fluctuation fixed point. It is obtained
for $\epsilon_d=0$ and $V_k=0$, $Upoint occurs in our case as well. But it is
unstable with respect to the hybridization $V_k$, and, by definiton,
unstable with respect to changes in $\epsilon_d$.
Since we introduced the flow equations to bring the Hamiltonian
closer to diagonalization, the case $V_k=0$ is trivial from our
point of view.

\section{Conclusions} \label{SC}
The aim of this paper was to show that Wegner's original flow equations
\cite{Wegner} together with a simple approximation yield
approximate flow equations that are simple to analyse. The
approximation consists of neglecting terms in the flow equation which
do not occur in the initial Hamiltonian. From a physical point of view
this may be reasonable, since all of these terms have a simple
physical meaning and should occur in a more realistic model. But
one does not expect a significantly different behaviour of the
system if one neglects these terms in the original Hamiltonian.
Therefore they might be irrelevant for the flow equations too. From
a mathematical point of view the approximation is not yet
understood. We only showed that for irrelevant terms our results are
equivalent to a second order perturbational treatment if the
off-diagonal matrix elements are small. But since $\eta\rightarrow 0$
as $\ell\rightarrow\infty$, all the matrix elements of $H$ are
functions of $\ell$ without any pole that tend to a certain value
for $\ell\rightarrow\infty$. Also the approximate flow equations have
this property. Therefore one should be able to estimate the error
made by the approximation. Further investigations in this direction
will be done.

On the other hand we are able
to discuss the relevance of the neglected terms a posteriori.
It is possible to introduce these terms in the flow equations
and to study the effectit cannot be simply included in $H^r$ since divergencies
occur. For example $V_k$
is a relevant contribution in the strong coupling fixed point.
If a contribution does only renormalize the other matrix elements
and vanishes for $\ell\rightarrow\infty$, it is irrelevant. If
it does not vanish it is marginal. In our example,
the Anderson model, we were able to show that the results obtained
in this way agree with results from a numerical renormalization group
approach \cite{KWW}. The main advantage of our approach is that
flow equations are obtained without any difficulty. One simply has
to calculate two commutators. Furthermore the flow equations
(\ref{flk}-\ref{flU}) may easily be generalized to the case
of many defects or to the case of degenerate defect states.
Especially the limit were the degeneracy is infinite has been
studied (see e.g. \cite{Hewson} and the references therein). This
limit can be studied using our approach as well and
it is possible to derive flow equations without any approximation.
We will use this limit in a subsequent paper to test our approximation.

\vskip1cm
The authors would like to thank F.~Wegner for many helpful discussions.
\vskip1cm

\begin{appendix}
\section{Derivation of the flow equations in section 4} \label{App}
In this appendix we derive the flow equations presented in section
\ref{SRio}. The new contribution is added to $H$ and $H^r$ and we calculate
the new $\eta=[H,H^r]$. Then the commutator $[\eta,H]$ contains several
new terms compared to (\ref{cetaH}). Some of these new terms are of a
form different from the contributions to $H$. These terms are neglected.
The other terms are calculated, since they of $H$ with respect to $\ell$. Then
the new flow equations are given.

\subsection{$\sum_{k,q,\sigma}V_{k,q}:c^{\dagg}_{k,\sigma}c_{q,\sigma}:$}
\beqar
H &=& \sum_{k,\sigma}\epsilon_k:c^{\dagg}_{k,\sigma}c_{k,\sigma}:
+\sum_{\sigma}\tilde{\epsilon}_d:d^{\dagg}_{\sigma}d_{\sigma}:
+\sum_{k,\sigma}V_k(:c^{\dagg}_{k,\sigma}d_{\sigma}:
+:d^{\dagg}_{\sigma}c_{k,\sigma}:)
\ret
&+& U:d^{\dagg}_{+}d^{\dagg}_{-}d_{-}d_{+}:
+\sum_{k,q,\sigma}V_{k,q}:c^{\dagg}_{k,\sigma}c_{q,\sigma}:.
\eeqar
\beq
H^r=\sum_{k,\sigma}V^r_k(:c^{\dagg}_{k,\sigma}d_{\sigma}:
+:d^{\dagg}_{\sigma}c_{k,\sigma}:)
+\sum_{k,q,\sigma}V_{k,q}:c^{\dagg}_{k,\sigma}c_{q,\sigma}:
\eeq
\beqar \label{Aeta1}
\eta &=& [H,H^r]
\ret
&=& \sum_{k,\sigma}\eta_k(:c^{\dagg}_{k,\sigma}d_{\sigma}:
-:d^{\dagg}_{\sigma}c_{k,\sigma}:)
+\sum_{k,q,\sigma}\eta_{k,q}(:c^{\dagg}_{k,\sigma}c_{q,\sigma}:
-:c^{\dagg}_{q,\sigma}c_{k,\sigma}:)
\ret
&+& \sum_{k,\sigma}\eta^{(2)}_k(:c^{\dagg}_{k,\sigma}
d^{\dagg}_{-\sigma}d_{-\sigma}d_{\sigma}:
-:d^{\dagg}_{\sigma}d^{\dagg}_{-\sigma}d_{-\sigma}c_{k,\sigma}:)
\eeqar
\beq
\eta_k = (\epsilon_k-\tilde{\epsilon}_d)V^r_k
+\sum_qV_{k,q}(V^r_q-V_q)
\eeq
\beq
\eta_{k,q} = \frac{1}{2}(\epsilon_k-\epsilon_q)V^r_{k,q}
+\frac{1}{2}(V_kV^r_q-V_qV^r_k)
\eeq
\beq
\eta^{(2)}_k = -U\,V^r_k.
\eeq
New contributions to $[\eta,H]$ are
\beqar
&-& \sum_{k,q,\sigma}\eta_qV_{k,q}
(:c^{\dagg}_{k,\sigma}d_{\sigma}:+:d^{\dagg}_{\sigma}c_{k,\sigma}:)
\ret
&+& 2\sum_{k,q,p,\sigma}(\eta_{k,p}V_{p,q}+\eta_{q,p}V_{p,k})
:c^{\dagg}_{k,\sigma}c_{q,\sigma}:
\ret
&-& \sum_{k,q,\sigma}\eta^{(2)}_qV_{k,q}
(:c^{\dagg}_{k,\sigma}d^{\dagg}_{-\sigma}d_{-\sigma}d_{\sigma}:
+:d^{\dagg}_{\sigma}d^{\dagg}_{-\sigma}d_{-\sigma}c_{k,\sigma}:)
\eeqar
The ldoes not contain a term of this form. It will be neglected. The first
term contributes to the derivative of $V_k$, the second term to the
derivative of $V_{k,q}$. Thus we obtain
\beq \label{Afli1k}
\frac{d\epsilon_k}{d\ell} = 2\eta_kV_k+4\sum_q\eta_{k,q}V_{k,q}
\eeq
\beq \label{Afli1d}
\frac{d\tilde{\epsilon}_d}{d\ell} = -2\sum_k\eta_kV_k
+2\sum_k\eta_k^{(2)}V_k(n_k-n_d)+U\frac{dn_d}{d\ell}
\eeq
\beq \label{Afli1V}
\frac{dV_k}{d\ell} = -\eta_k(\epsilon_k-\tilde{\epsilon}_d)
+2\sum_p\eta_{k,p}V_p+n_d(1-n_d)\eta_k^{(2)}U-\sum_p\eta_pV_{k,p}
\eeq
\beq \label{Afli1Vkq}
\frac{dV_{k,q}}{d\ell} = \eta_kV_q+\eta_qV_k
-\eta_{k,q}(\epsilon_k-\epsilon_q)
+2\sum_p(\eta_{k,p}V_{p,q}+\eta_{q,p}V_{p,k})
\eeq
\beq \label{Afli1U}
\frac{dU}{d\ell}=-4\sum_k\eta_k^{(2)}V_k
\eeq
Using the expression for the different matrix elements of $\eta$ given
above we obtain (\ref{fli1k}-\ref{fli1U})

\subsection{$\sum_{k,\sigma}V_k^{(2)}
(:c^{\dagg}_{k,\sigma}d^{\dagg}_{-\sigma}d_{-\sigma}d_{\sigma}:
+:d^{\dagg}_{\sigma}d^{\dagg}_{-\sigma}d_{-\sigma}c_{k,\sigma}:)$}
\beqar
H &=& \sum_{k,\sigma}\epsilon_k:c^{\dagg}_{k,\sigma}c_{k,\sigma}:
+\sum_{\sigma}\tilde{\epsilon}_d:d^{\dagg}_{\sigma}d_{\sigma}:
+\sum_{k,\sigma}V_k(:c^{\dagg}_{k,\sigma}d_{\sigma}:
+:d^{\dagg}_{\sigma}c_{k,\sigma}:)
\ret
&+& U:d^{\dagg}_{+}d^{\dagg}_{-}d_{-}d_{+}:
+\sum_{k,\sigma}V_k^{(2)}
(:c^{\dagg}_{k,\sigma}d^{\dagg}_{-\sigma}d_{-\sigma}d_{\sigma}:
+:d^{\dagg}_{\sigma}d^{\dagg}_{-\sigma}d_{-\sigma}c_{k,\sigma}:)
\eeqar
\beq
H^r=\sum_{k,\sigma}V^r_k(:c^{\dagg}_{k,\sigma}d_{\sigma}:
+:d^{\dagg}_{\sigma}c_{k,\sigma}:).
+\sum_{k,\sigma}V_k^{(2)}
(:c^{\dagg}_{k,\sigma}d^{\dagg}_{-\sigma}d_{-\sigma}d_{\sigma}:
+:d^{\dagg\eeq
\beqar \label{eta3}
\eta &=& [H,H^r]
\ret
&=& \sum_{k,\sigma}\eta_k(:c^{\dagg}_{k,\sigma}d_{\sigma}:
-:d^{\dagg}_{\sigma}c_{k,\sigma}:)
+\sum_{k,q,\sigma}\eta_{k,q}(:c^{\dagg}_{k,\sigma}c_{q,\sigma}:
-:c^{\dagg}_{q,\sigma}c_{k,\sigma}:)
\ret
&+& \sum_{k,\sigma}\eta^{(2)}_k(:c^{\dagg}_{k,\sigma}
d^{\dagg}_{-\sigma}d_{-\sigma}d_{\sigma}:
-:d^{\dagg}_{\sigma}d^{\dagg}_{-\sigma}d_{-\sigma}c_{k,\sigma}:)
\ret
&+& \sum_{k,q,\sigma}\eta^{(2)}_{k,q}
(:c^{\dagg}_{k,\sigma}d^{\dagg}_{-\sigma}d_{-\sigma}c_{q,\sigma}:
-:c^{\dagg}_{q,\sigma}d^{\dagg}_{-\sigma}d_{-\sigma}c_{k,\sigma}:
\ret & & \qquad
+:d^{\dagg}_{\sigma}c^{\dagg}_{k,-\sigma}d_{-\sigma}c_{q,\sigma}:
-:c^{\dagg}_{q,\sigma}d^{\dagg}_{-\sigma}c_{k,-\sigma}d_{\sigma}:
\ret & & \qquad
+:c^{\dagg}_{q,\sigma}c^{\dagg}_{k,-\sigma}d_{-\sigma}d_{\sigma}:
-:d^{\dagg}_{\sigma}d^{\dagg}_{-\sigma}c_{k,-\sigma}c_{q,\sigma}:)
\eeqar
\beq
\eta_k = (\epsilon_k-\tilde{\epsilon}_d)V^r_k-n_d(1-n_d)U\,V^{(2)}_k
\eeq

\beq
\eta_{k,q} = \frac{1}{2}(V_kV^r_q-V_qV^r_k)
\eeq

\beq
\eta^{(2)}_k = -U\,V^r_k
+(\epsilon_k-\tilde{\epsilon}_d-(1-2n_d)U)V^{(2)}_k
\eeq

\beq
\eta^{(2)}_{k,q} = \frac{1}{2}((V_k-V^r_k)V^{(2)}_q-(V_q-V^r_q)V^{(2)}_k)
\eeq
The following new terms in $[\eta,H]$ (compared to (\ref{cetaH}))
contribute to the derivative of $H$ with respect to $\ell$:
\beqar
&-&2\sum_{k,\sigma}(\eta_k V_k^{(2)}+(1-2n_d)\eta_k^{(2)} V_k^{(2)})
:d^{\dagg}_{\sigma}d^{\dagg}_{-\sigma}d_{-\sigma}d_{\sigma}:
\ret
&+&2\sum_{k,\sigma}n_d(1-n_d)\eta_k^{(2)} V_k^{(2)}
:c^{\dagg}_{k,\sigma}c_{k,\sigma}:
\ret
&+&2\sum_{k,\sigma}((n_k-n_d)\eta_k V_k^{(2)}
-n_d(1-n_d)\eta_k^{(2)} V_k^{(2)})
:d^{\dagg}_{\sigma}d_{\sigma}:
\ret
&+&2\sum_{k,q,\sigma}(\eta_{k,q}^{(2)}V_q+\eta_{k,q}V_q^{(2)}
+(1-2n_d)\eta_{k,q}^{(2)}V_q^{(2)})
(:c^{\dagg}_{k,\sigma}d^{\dagg}_{-\sigma}d_{-\sigma}d_{\sigma}:
+:d^{\dagg}_{\sigma}d^{\dagg}_{-\sigma}d_{-\sigma}c_{k,\sigma}:)
\ret
&+&2\sum_{k,q,\sigma}\eta_{k,q}^{(2)}V_q^{(2)}n_d(1-n_d)
(:c^{\dagg}_{k,\sigma}d_{\sigma}:
+:d^{\dagg}_{\sigma}c_{k,\sigma}:)
\eeqar
The flow equations are now
\beq \label{Afli2k}
\frac{d\epsilon_k}{d\ell}
=2\eta_kV_k+2n_d(1-n_d)\eta_k^{(2)} V_k^{(2)}
\eeq
\beq \label{Afli2d}
\frac{d\tilde{\epsilon}_d}{d\ell} = -2\sum_k\eta_kV_k
+2\sum_k \left((n_k-n_d)(\eta_k^{(2)}V_k+\eta_k V_k^{(2)})
-n_d(1-n_d)\eta_k^{(2)} V_k^{(2)}\right)
+U\frac{dn_d}{d\ell}
\eeq
\beq \label{Afli2Vk}
\frac{dV_k}{d\ell} =-\eta_k(\epsilon_k-\tilde{\epsilon}_d)
+2\sum_p(\eta_{k,p}V_p+\eta_{k,p}^{(2)}V_p^{(2)}n_d(1-n_d))
+Un_d(1-n_d)\eta_k^{(2)}
\eeq
\beq \label{Afli2U}
\frac{dU}{d\ell}=-4\sum_k(\eta_k^{(2)}V_k
+\eta_k V_k^{(2)}+(1-2n_d)\eta_k^{(2)} V_k^{(2)})
\eeq
\beq \label{Afli2V2}
\frac{dV_k^{(2)}}{d\ell} =
\eta_kU-\eta_k^{(2)}(\epsilon_k-\tilde{\epsilon}_d-(1-2n_d)U)
+2\sum_q(\eta_{k,q}^{(2)}V_q+\eta_{k,q}V_q^{(2)}
+(1-2n_d)\eta_{k,q}^{(2)}V_q^{(2)})
\eeq
Using the matrix elements of $\eta$ calculated above we obtain
(\ref{fli2k}-\ref{fli2V2}).

\subsection{$
\sum_{k,q,\sigma}V_{k,q}^{(2)}
(:c^{\dagg}_{k,\sigma}d^{\dagg}_{-\sigma}d_{-\sigma}c_{q,\sigma}:
-:c^{\dagg}_{k,\sigma}d^{\dagg}_{-\sigma}d_{\sigma}c_{q,-\sigma}:).
$}
$H$ and $H^r$ are obtained as before adding this term. This yields
\beqar \label{eta4}
\eta &=& \sum_{k,\sigma}\eta_k(:c^{\dagg}_{k,\sigma}d_{\sigma}:
-:d^{\dagg}_{\sigma}c_{k,\sigma}:)
+\sum_{k,q,\sigma}\eta_{k,q}(:c^{\dagg}_{k,\sigma}c_{q,\sigma}:
-:c^{\dagg}_{q,\sigma}c_{k,\sigma}:)
\ret
&+& \sum_{k,\sigma}\eta^{(2)}_k(:c^{\dagg}_{k,\sigma}
d^{\dagg}_{-\sigma}d_{-\sigma}d_{\sigma}:
-:d\ret
&+& \sum_{k,q,\sigma}\eta^{(2)}_{k,q}
(:c^{\dagg}_{k,\sigma}d^{\dagg}_{-\sigma}d_{-\sigma}c_{q,\sigma}:
-:c^{\dagg}_{q,\sigma}d^{\dagg}_{-\sigma}d_{-\sigma}c_{k,\sigma}:
\ret & & \qquad
-:c^{\dagg}_{k,\sigma}d^{\dagg}_{-\sigma}d_{\sigma}c_{q,-\sigma}:
+:c^{\dagg}_{q,\sigma}d^{\dagg}_{-\sigma}d_{\sigma}c_{k,-\sigma}:)
\ret
&+& \sum_{k,q,\sigma}\eta^{(2)}_{k,q,p}
(:c^{\dagg}_{k,\sigma}c^{\dagg}_{p,-\sigma}d_{-\sigma}c_{q,\sigma}:
-:c^{\dagg}_{q,\sigma}d^{\dagg}_{-\sigma}c_{p,-\sigma}c_{k,\sigma}:
\ret & & \qquad
+:c^{\dagg}_{k,\sigma}d^{\dagg}_{-\sigma}c_{p,-\sigma}c_{q,\sigma}:
-:c^{\dagg}_{q,\sigma}c^{\dagg}_{p,-\sigma}d_{-\sigma}c_{k,\sigma}:
\ret & & \qquad
-:c^{\dagg}_{k,\sigma}c^{\dagg}_{p,-\sigma}d_{\sigma}c_{q,-\sigma}:
+:c^{\dagg}_{q,\sigma}d^{\dagg}_{-\sigma}c_{p,\sigma}c_{k,-\sigma}:
\ret & & \qquad
-:c^{\dagg}_{k,\sigma}d^{\dagg}_{-\sigma}c_{p,\sigma}c_{q,-\sigma}:
+:c^{\dagg}_{q,\sigma}c^{\dagg}_{p,-\sigma}d_{\sigma}c_{k,-\sigma}:)
\eeqar
\beq
\eta_k = (\epsilon_k-\tilde{\epsilon}_d)V^r_k
-\sum_qV_{k,q}^{(2)}(V^r_q-V_q)(n_q-n_d)
\eeq

\beq
\eta_{k,q} = \frac{1}{2}(V_kV^r_q-V_qV^r_k)
\eeq

\beq
\eta^{(2)}_k = -U\,V^r_k
+\sum_qV_{k,q}^{(2)}(V^r_q-V_q)
\eeq

\beq
\eta^{(2)}_{k,q} = \frac{1}{2}(\epsilon_k-\epsilon_q)V_{k,q}^{(2)}
\eeq

\beq
\eta^{(2)}_{k,q,p} = \frac{1}{2}V_{k,q}^{(2)}(V_p-V^r_p)
\eeq
The following additional terms in $[\eta,H]$ (compared to (\ref{cetaH}))
are of the same form as terms in $H$ and have to be taken into account.
\beqar
&-&\sum_{k,q,\sigma}(\eta_qV_{k,q}^{(2)}-2\eta_{k,q}^{(2)}V_q)(n_q-n_d)
(:c^{\dagg}_{k,\sigma}d_{\sigma}:+:d^{\dagg}_{\sigma}c_{k,\sigma}:)
\ret
&+&2\sum_{k,q,p,\sigma}(\eta_{k,p}V_{p,q}^{(2)}+\eta_{q,p}V_{p,k}^{(2)})
(:c^{\dagg}_{k,\sigma}d^{\dagg}_{-\sigma}d_{-\sigma}c_{q,\sig-:c^{\dagg}_{k,\sigma}d^{\dagg}_{-\sigma}d_{\sigma}c_{q,-\sigma}:)
\ret
&-&2\sum_{k,q,\sigma}\eta_{k,q}V_{k,q}^{(2)}(n_k-n_q)
:d^{\dagg}_{\sigma}d_{\sigma}:
\ret
&-&2n_d(1-n_d)\sum_{k,q,\sigma}\eta_q^{(2)}V_{k,q}^{(2)}
(:c^{\dagg}_{k,\sigma}d_{\sigma}:+:d^{\dagg}_{\sigma}c_{k,\sigma}:)
\ret
&+&4(1-2n_d)\sum_{k,q,p,\sigma}
(\eta_{k,p}^{(2)}V_{p,q}^{(2)}+\eta_{q,p}^{(2)}V_{p,k}^{(2)})
(:c^{\dagg}_{k,\sigma}d^{\dagg}_{-\sigma}d_{-\sigma}c_{q,\sigma}:
-:c^{\dagg}_{k,\sigma}d^{\dagg}_{-\sigma}d_{\sigma}c_{q,-\sigma}:)
\ret
&-&4(1-2n_d)\sum_{k,q,\sigma}\eta_{k,q}^{(2)}V_{k,q}^{(2)}(n_k-n_q)
:d^{\dagg}_{\sigma}d_{\sigma}:
\ret
&-&2\sum_{k,q,\sigma}\eta_{k,q}^{(2)}V_{k,q}^{(2)}(n_k-n_q)
:d^{\dagg}_{\sigma}d^{\dagg}_{-\sigma}d_{-\sigma}d_{\sigma}:
\ret
&+&4\sum_{k,q,p,\sigma}\eta_{p,q,k}^{(2)}V_{p,q}^{(2)}
[n_p(1-n_q)(1-n_d)-(1-n_p)n_qn_d]
(:c^{\dagg}_{k,\sigma}d_{\sigma}:+:d^{\dagg}_{\sigma}c_{k,\sigma}:)
\ret
&-&2\sum_{k,q,\sigma}\eta_{k,q}^{(2)}(\epsilon_k-\epsilon_q)
(:c^{\dagg}_{k,\sigma}d^{\dagg}_{-\sigma}d_{-\sigma}c_{q,\sigma}:
-:c^{\dagg}_{k,\sigma}d^{\dagg}_{-\sigma}d_{\sigma}c_{q,-\sigma}:)
\ret
&-&4\sum_{k,q,p,\sigma}\eta_{k,q,p}^{(2)}V_p
(:c^{\dagg}_{k,\sigma}d^{\dagg}_{-\sigma}d_{-\sigma}c_{q,\sigma}:
-:c^{\dagg}_{k,\sigma}d^{\dagg}_{-\sigma}d_{\sigma}c_{q,-\sigma}:)
\eeqar
Now the flow equations are
\beq \label{Afli4k}
\frac{d\epsilon_k}{d\ell}=2\eta_kV_k
\eeq

\beqar \label{Afli4d}
\frac{d\tilde{\epsilon}_d}{d\ell} &=&
-2\sum_k\eta_kV_k
+2\sum_k\eta_k^{(2)}V_k(n_k-n_d)+U\frac{dn_d}{d\ell}
\ret & &
-2\sum_{k,q}\eta_{k,q}V_{k,q}^{(2)}(n_k-n_q)
-4(1-2n_d)\sum_{k,q,\sigma}\eta_{k,q}^{(2)}V_{k,q}^{(2)}(n_k-n_q)
\eeqar

\beqar \label{Afli4Vk}
\frac{dV_k}{d\ell} &=&-\eta_k(\epsilon_k-\tilde{\epsilon}_d)
+2\sum_q\eta_{k,q}V_q+Un_d(1-n_d)\eta_k^{(2)}
\ret & &
-\sum_q(\eta_qV_{k,q}^{(2)}-\eta_{k,-2n_d(1-n_d)\sum_q\eta_q^{(2)}V_{k,q}^{(2)}
\ret & &
+4\sum_{q,p}\eta_{p,q,k}^{(2)}V_{p,q}^{(2)}[n_p(1-n_q)(1-n_d)-(1-n_p)n_qn_d]
\eeqar

\beq \label{Afli4U}
\frac{dU}{d\ell}=-4\sum_k\eta_k^{(2)}V_k
-2\sum_{k,q}\eta_{k,q}^{(2)}V_{k,q}^{(2)}(n_k-n_q)
\eeq

\beqar \label{Afli4V2}
\frac{dV_{k,q}^{(2)}}{d\ell}&=&\eta^{(2)}_kV_q+\eta^{(2)}_qV_k
-2\eta_{k,q}^{(2)}(\epsilon_k-\epsilon_q)
-4\sum_p\eta_{k,q,p}^{(2)}V_p
\ret & &
+2\sum_p(\eta_{k,p}V_{p,q}^{(2)}+\eta_{q,p}V_{p,k}^{(2)})
+4(1-2n_d)\sum_p(\eta_{k,p}^{(2)}V_{p,q}^{(2)}+\eta_{q,p}^{(2)}V_{p,k}^{(2)})
\eeqar
As before we obtain (\ref{fli4k}-\ref{fli4V2}).

\end{appendix}



\begin{thebibliography}{99}
\bibitem{Wegner} Wegner, F.: Ann. Physik {\bf 3}, 77--91 (1994)
\bibitem{SW} Schrieffer, J.R., and Wolff, P.A.: Phys. Rev {\bf 149}, 491--492
(1966)
\bibitem{Anderson} Anderson, P.W.: Phys. Rev {\bf 124}, 41--53 (1961)
\bibitem{Hewson} Hewson, A.C.: {\em The Kondo Problem to Heavy Fermions.\/}
Cambridge Studies in Magnetism Vol. 2 (Cambridge University Press,
Cambridge, 1993).
\bibitem{KWW} Krishnamurty, H.R., Wilkins, and J.W., Wilson, K.G.:
Phys. Rev. {\bf B21}, 1003--1043 and 1044--1083 (1980).
\end{thebibliography}
\end{document}